\documentclass[twocolumn,showpacs,preprintnumbers,amsmath,amssymb]{revtex4}%
\usepackage{graphicx} 
\usepackage{dcolumn} 
\usepackage{bm} 
\usepackage{amsmath}
\usepackage{amsfonts}
\usepackage{amssymb}

\begin{document} 
 
\title{Internal state conversion in ultracold gases} 
\author{J.N.\ Fuchs\footnote{email: fuchs@lkb.ens.fr}, D.M.\
  Gangardt$^{||,\footnote{email: gangardt@lkb.ens.fr}}$ and F.\ Lalo\"{e}\footnote{email: laloe@lkb.ens.fr}} 
\affiliation{LKB, Laboratoire de Physique de l'ENS, 24 rue Lhomond, 75005 Paris, France\\ 
$(^{||})$Also at Department of physics, Technion, 32000 Haifa, Israel} 
 
\date{\today} 
 
\begin{abstract} 
We consider an ultracold gas of (non-condensed) bosons or fermions with two 
internal states, and study the effect of a gradient of the transition 
frequency between these states. When a $\pi/2$ RF pulse is applied to the 
sample, exchange effects during  collisions transfer the atoms into internal states 
which depend on the direction of their velocity. This results, after a short 
time, in a spatial separation between the two states. A kinetic equation is 
solved analytically and numerically; the results agree well with the recent 
observations of Lewandowski et al. 
 
\end{abstract} 
 
\maketitle 
 
 
In the last few years, the study of ultracold gases has generated a wealth of 
very interesting results.\ Spectacular examples are given by Bose 
condensed gases, but gases above their degeneracy temperature also  
provide exciting and unexpected results.\ For instance, recent experiments by 
Lewandowski et al. \cite{Cornell} have shown the existence of a remarkable 
phenomenon, observed when a RF pulse is applied to a $^{87}$Rb gas with 
two internal states, cooled by laser irradiation and evaporative cooling (but 
not Bose condensed).\ Since the two internal states are similar to two 
different species of atoms, the authors describe their observation as a 
``segregation'' between the species.\ They also mention that the differential Stern-Gerlach force, 
due to the magnetic gradient acting on the species, is too small to explain 
the segregation, which is actually related to interactions between the atoms. 
The purpose of the present article is to show that the ``identical spin 
rotation effect'' (ISRE) provides a qualitative and quantitative explanation 
of the observations. 
 
The ISRE was introduced in \ \cite{LL1} as a microscopic phenomenon taking 
place during a binary collision between two identical atoms with internal 
degrees of freedom, nuclear spins for instance.\ The effect is a consequence 
of quantum indistinguishability; it introduces a rotation of each spin around 
their sum (in opposite directions for bosons and fermions).\ For instance, if 
a single atom with a spin polarization in a given direction crosses a
gas of identical atoms polarized in another direction, the spin
of the transmitted atom undergoes a rotation; 
this is similar to the rotation of the polarization of photons in the Faraday 
effect.\ On a macroscopic scale, the effect can affect transport properties of 
gases with internal states.\ For instance, ref.\ \cite{LL2} considers a gas 
which is in a ``classical'' regime in terms of equilibrium properties, but 
where quantum effects are important in binary collisions.\ When the density is 
sufficient to reach an hydrodynamic regime, this work shows the existence of 
transverse spin waves, analogous to spin waves in degenerate liquid $^{3}$He 
\cite{Leggett}. Similar predictions had been made independently by Bashkin 
\cite{Bashkin} from a more macroscopic point of view, based on the notion of 
``molecular field'' (or mean field) - see also the work of L\'{e}vy and 
Ruckenstein \cite{Levy-Ruckenstein}. Transverse spin waves in gases were 
subsequently observed in H$\downarrow$ \cite{Johnson} as well as in helium 
\cite{helium,Gully-Mullin}. 
 
Another prediction made in ref.\ \cite{LL2} (end of section 1) is that the 
ISRE can also create ``longitudinal oscillations'' when a $\pi/2$ pulse is 
applied to the sample, provided the transverse spin polarization is 
inhomogeneous.\ Here we show that the phenomenon described in ref. 
\cite{Cornell} is precisely this effect, transposed to the pseudo spin 
associated with the two hyperfine levels relevant in the experiment, as 
foreseen by the authors who mention a ``longitudinal spin effect'' in their 
conclusion. The major differences are that the experiment was performed at a 
density where the hydrodynamic regime is not reached, and that the spin 
oscillations are not of small amplitude. 
 
A point which emerged from the early studies on spin waves in gases, sometimes 
after vivid controversy, is that the effect of binary collisions in a gas are 
well described by a simple mean field calculation, provided one
considers forward scattering only.\ In condensed matter, each particle interacts at 
the same time with several others; it seems natural that their individual 
effects should be well averaged by the test particle so that mean field theory 
should apply.\ By contrast, in a dilute gas, particles are ``free almost all 
the time''; they interact only during brief collisions, with a single partner 
with which they can develop strong correlations.\ Indeed, in atomic physics, 
one rarely studies collision processes within mean field 
theory!\ Nevertheless, it turns out that the average effect of many collisions 
in the forward direction is equivalent to the results of mean field
theory, if one replaces the 
real binary interaction potential by a pseudopotential involving directly the 
scattering length (using the real potential would lead to meaningless 
results); the equivalence holds in the limit of low collision energy (the ISRE in the forward
 direction dominates over 
lateral scattering at very low energies since the corresponding ``cross 
section'' $\tau_{ex}^{fwd.}$ diverges \cite{LL1}). 
 
In the experimental conditions of ref.\ \cite{Cornell}, the atoms are in
an axially symmetric magnetic trap elongated in the $Ox$ 
direction. Initially the gas is at equilibrium with only state $1$ populated. \ One 
then applies a $\pi/2$ RF pulse which, suddenly, puts all the atoms into the 
same coherent superposition of states $1$ and $2$, corresponding to a uniform 
transverse polarization of the pseudo spin.\ The system is then left free to 
evolve, and one observes the time evolution of the local densities $n_{1}$ and 
$n_{2}$. 

We begin with a qualitative physical discussion of the sequence 
of events.\ Since the field gradient creates an inhomogeneous spin precession, the gas develops a 
gradient of transverse spin orientation: correlations are created between 
position and transverse spin orientation. The free thermal motion of 
the atoms then creates correlations between velocity and transverse spin. 
Thus, a particle moving with a given velocity at point $x$ gets a spin 
polarization which is not parallel to the average local spin polarization, so 
that the ISRE precession takes place.\ This makes its spin polarization leave 
the transverse plane and develop a non-zero value of its longitudinal  component, with an 
opposite sign for different signs of the $x$-component of the velocity of the 
atom. The appearance of this component indicates the beginning of an internal 
conversion, which eventually results in spatial separation of the atoms in 
different internal states. We emphasize that the apparent segregation is not 
the result of a spatial separation of atoms in fixed internal states, as for 
two different chemical species; on the contrary, without changing their 
spatial position, the ISRE transfers atoms into internal states that 
depend on their motion. 
 
For a more quantitative discussion, we use a transport equation in terms of a 
time $t$ dependent operator $\widehat{\rho}(\mathbf{r},\mathbf{p},t)$, which 
depends on position $\mathbf{r}$ and momentum $\mathbf{p}$; $\widehat{\rho}$ 
is the Wigner transform with respect to orbital variables of the single 
particle density matrix; it remains a $2\times2$ operator in the space of 
internal variables, corresponding to states $1$ and $2$. Instead of
using the 4 matrix elements of $\widehat{\rho}$, it is often 
convenient to replace them by a local density $f$ and (pseudo) spin 
density $\mathbf{M}$ in phase space defined by: 
\begin{equation} 
\widehat{\rho}(\mathbf{r},\mathbf{p},t)=\frac{1}{2}\left[  f(\mathbf{r}%
,\mathbf{p},t)\widehat{I}+\mathbf{M}(\mathbf{r},\mathbf{p},t)\cdot 
\widehat{\mathbf{\sigma}}\right]  ,\label{rho}%
\end{equation} 
where $\widehat{I}$ is the unit operator in spin space and $\widehat 
{\mathbf{\sigma}}$ the spin operator whose three components are the Pauli 
matrices.\ The kinetic equation for $\widehat{\rho}(\mathbf{r},\mathbf{p},t)$
(see for instance \cite{LL1}) is: 
\begin{align} 
\partial_{t}\widehat{\rho} &  +\frac{\mathbf{p}}{m}\cdot\mathbf{\nabla 
}_{\mathbf{r}}\widehat{\rho}-\frac{1}{2}\left[  \mathbf{\nabla}_{\mathbf{p}%
}\widehat{\rho},\cdot\mathbf{\nabla}_{\mathbf{r}}\widehat{U}(\mathbf{r}%
,t)\right]  _{+}+\frac{1}{i\hbar}\left[  \widehat{\rho},\widehat{U}%
(\mathbf{r},t)\right]  _{-}\nonumber\\ 
&  =I_{coll}[\widehat{\rho}],\label{kinetic}%
\end{align} 
where the second term is the usual drift term ($m$ is the mass of the 
particles); the third term (anticommutator) is the force term including both 
the effect of the trapping potential and of the mean field created by the 
other atoms; the fourth term (commutator) is a spin precession term containing 
the ISRE as well as some other contributions that we discuss below - this 
commutator is the term on which we focus our attention in this article.\ In 
the right hand side, the collision integral $I_{coll}[\widehat{\rho}]$ 
describes ``real'' collisions (lateral scattering as opposed to forward 
scattering, already included in the mean field); it can be obtained for 
instance from the LL transport equation \cite{LL1}, or even take a more 
detailed expression containing ``non-local collision terms'' with $\mathbf{r}$ 
and $\mathbf{p}$ gradients, as discussed e.g. in the appendix of \cite{Snider} 
and \cite{Meyerovich}. In fact, if we are mostly interested in a Knudsen regime, the 
precise expression of $I_{coll}[\widehat{\rho}]$ is not needed.\ The effective 
potential $\widehat{U}(\mathbf{r},t)$ is the spin operator: 
\begin{equation} 
\widehat{U}(\mathbf{r},t)=U_{0}(\mathbf{r},t)\widehat{I}+\mathbf{U}%
(\mathbf{r},t)\cdot\widehat{\mathbf{\sigma}},\label{U}%
\end{equation} 
where the scalar component is defined by:%
\begin{equation} 
U_{0}=\frac{V_{1}+V_{2}}{2}+g_{22}^{\epsilon}\,n_{2}+g_{11}^{\epsilon}%
\,n_{1}+\frac{g_{12}^{\epsilon}}{2}(n_{2}+n_{1}).\label{scalar}%
\end{equation} 
Here $V_{1}$ and $V_{2}$ are the external trapping potentials acting on states 
$1$ and $2$; the $g^{\epsilon}$'s have the following expressions in terms of 
the usual ``coupling constants'' $g$, proportional to the appropriate 
scattering lengths associated with the various possibilities for pair 
interactions between atoms in levels $1$ or $2$ \cite{elastic}: 
\begin{equation} 
g_{11,22}^{\epsilon}=g_{11,22}(1+\epsilon)/2\,\ \text{;}\,\,g_{12}^{\epsilon 
}=g_{d}+\epsilon g_{t}\,\,,\label{g's}%
\end{equation} 
where $g_{d}$ and $g_{t}$ refer to the direct and transfer process for two 
atoms in different levels. The number densities of atoms in levels $1$ and $2$ 
are $n_{1,2}$; $\epsilon=+1$ ($-1$) for bosons (fermions).\ The vectorial 
component of $\widehat{U}(\mathbf{r},t)$ is:%
\begin{equation} 
\mathbf{U}(\mathbf{r},t)=\frac{\hbar\Omega(\mathbf{r},t)}{2}\mathbf{e}_\| 
+\epsilon\frac{g_{12}^{\epsilon}}{2}\mathbf{m}(\mathbf{r}%
,t);\label{vectorial}%
\end{equation} 
$\mathbf{e}_{\|}$ is the unit vector in the longitudinal spin direction, and 
$\Omega(\mathbf{r},t)$ is: 
\begin{equation} 
\hbar\Omega=V_{2}-V_{1}+2g_{22}^{\epsilon}n_{2}-2g_{11}^{\epsilon}%
n_{1}+2g_{12}^{\epsilon}(n_{1}-n_{2}),\label{omega}%
\end{equation} 
where the total density $n$ and spin polarization $\mathbf{m}$ are:%
\begin{equation} 
n(\mathbf{r},t)=\int d^{3}p\,\,f(\mathbf{r},\mathbf{p},t)\text{ ;\ }%
\mathbf{m}(\mathbf{r},t)=\int d^{3}p\;\mathbf{M}(\mathbf{r},\mathbf{p}%
,t)\label{nandm}%
\end{equation} 
and $n_{1,2}=(n\mp m_{\|})/2$. The 
first contribution (\ref{omega}) to $\mathbf{U}$ acts as a ``local magnetic 
field''; its average value over the sample can be removed in a 
uniformly rotating frame. The second contribution originates from the ISRE, 
and is proportional to the local spin polarization $\mathbf{m}$ (only
$\mathbf{m}$ enters the ISRE commutator because  the $1/k$ divergence 
of $\tau_{ex}^{fwd.}$ at low $k$'s \cite{LL1} compensates the relative 
velocity factor of the collision integral). The commutator makes $\mathbf{M}$ 
precess around the momentum integrated local spin polarization; 
it does not affect the evolution of $\mathbf{m}$ itself, but can change the 
evolution of $\mathbf{M}$ for each value of $\mathbf{p}$.

A few simplifying assumptions are appropriate in the experimental conditions 
of \cite{Cornell}. The confining energy is of order $k_B
T\simeq13$~kHz$\times h$ which is much 
larger than the mean-field interaction energy $gn(0)\simeq140$~Hz$\times h$
and the differential trapping energy $V_1-V_2\sim10$~Hz$\times h$. 
In the anti-commutator this allows us to retain only the confining energy of
the harmonic trap:
\begin{equation} 
U_{0}\simeq\frac{V_{1}+V_{2}}{2}=\frac{1}{2}m\left[  \omega^{2}x^{2}%
+\omega_{rad}^{2}(y^{2}+z^{2})\right]  ;\label{Uzero}%
\end{equation}
here $\omega$ and $\omega_{rad}$ are the axial 
and radial trapping frequencies. 
In the commutator of (\ref{kinetic}), $U_{0}$ 
disappears, and only the vectorial component $\mathbf{U}$ plays a role. 
The cigar shaped trap has an axial frequency $\omega/2\pi=7$~Hz, much smaller
than the radial frequency $\omega_{rad}/2\pi=230$~Hz, so that the system is
quasi one dimensional along the $x$ axis. Assuming that radial local equilibrium is quickly 
established in the $yz$ plane, we introduce on-axis value $\widehat{\rho}(x,p,t)$ (integrated over 
radial momenta). When averaged over radial coordinates and 
momenta, equation (\ref{vectorial}) becomes 
\begin{equation}
\overline{\mathbf{U}} 
(x,t)=\hbar\overline{\Omega}(x)\mathbf{e}_{\|}/2+\epsilon g_{12}^{\epsilon 
}\mathbf{m}(x,t)/4
\label{ubar}
\end{equation} 
where $\hbar\overline{\Omega}= 
\overline{V}_{2}-\overline{V}_{1}+(g_{22}^{\epsilon}-g_{11}^{\epsilon}) n/2$; 
note that the coupling 
constants are renormalized by a factor $1/2$  upon averaging
\cite{Levitov}; we have  assumed that $2g_{12}^{\epsilon}\simeq g_{11}^{\epsilon}+g_{22}^{\epsilon}$. 

With these assumptions the initial equilibrium Maxwell-Boltzmann 
distribution $f(x ,p)$ solves the kinetic equation
(\ref{kinetic}), so that the dynamics after the pulse can be expressed in terms of 
$\mathbf{M}$ only.  When lengths are 
measured in units of $x_{T}=\sqrt{k_{B}T/m\omega^{2}}$, momenta in units of 
$p_{T}=\sqrt{mk_{B}T}$ and times in units of $1/\omega$, equation 
(\ref{kinetic}) simplifies into: 
\begin{equation} 
\partial_{t}\mathbf{M}+p\partial_{x}\mathbf{M}-x\partial_{p}\mathbf{M}%
-\frac{2\overline{\mathbf{U}}}{\hbar\omega}\times\mathbf{M}\simeq 
-\frac{\mathbf{M}-\mathbf{M}^{eq}}{\omega \tau_{M}}\label{msimplifie}%
\end{equation} 
where a simple relaxation-time approximation has been made for $I_{coll}$ with 
a single parameter $\tau_{M}$ of the order of the time between collisions; the 
local equilibrium value $\mathbf{M}^{eq}=\mathbf{m}(x,t)\exp{(-p^{2}/2)} 
/\sqrt{2\pi}$. This is the equation that we now discuss.

For small times we can use a time expansion $\mathbf{M}(x,p,t)=\mathbf{M}%
^{(0)}+\mathbf{M}^{(1)}t+\mathbf{M}^{(2)}t^{2}/2+..$ and solve 
(\ref{msimplifie}) to each time order. The spin distribution 
immediately after the $\pi/2$ pulse is unchanged, except that 
$\mathbf{M}^{(0)}(x,p)$ is perpendicular to $\mathbf{e}_{\|}$. The density 
profile $n(x)$ remains Gaussian, so that the Bohr frequency $\overline{\Omega 
}(x)$ does not vary in time. The effect of ``real'' collisions (right-hand 
side of (\ref{msimplifie})) is neglected since we are interested in small time 
behavior only. The result of this calculation is that $m_{\|}(x,t)$ starts as 
$t^{4}$ with $m_{\|}^{(4)}(x)=\epsilon g_{12}^{\epsilon}n(x)\left[ 
\overline{\Omega}^{\prime\prime}(x)n(x)+\overline{\Omega}^{\prime 
}(x)\{n^{\prime}(x)-xn(x)\}\right]  /2\hbar$ in dimensionless form. Using the 
fact that the density $n(x)$ remains Gaussian and restoring the units, we 
get: 
\begin{equation} 
\frac{m_{\|}(x,t)}{n(x)}=\frac{n_{2}-n_{1}}{n}=\frac{\epsilon g_{12}^{\epsilon 
}\,n}{\hbar\omega}\,\frac{\overline{\Omega}^{\prime\prime}x_{T}^{2}%
-2\overline{\Omega}^{\prime}x}{\omega}\,\frac{(\omega t)^{4}}{48}%
.\label{exactresult}%
\end{equation} 
The first factor in the right hand side is of the order of  the dimensionless ISRE constant 
$ g_{12} n(0)/\hbar\omega$.
Since its value in the 
experiment of ref.~\cite{Cornell} is $\sim20$, it is not surprising that 
species separation could be observed in a time smaller than the trap period. 
The second factor involves the spatial variations of $\overline{\Omega}(x)$; 
near the center of the trap, a positive curvature implies a positive $m_{\|}$ 
($n_{2}\geq n_{1}$), in accordance with the results of \cite{Cornell} 
($\tau_{ex}^{fwd.}<0$). Inserting the values of the parameters of the 
experiment of \cite{Cornell} into (\ref{exactresult}) leads to significant 
species separation ($m_{\|}\sim n$) for $\sim25$~ms, to be compared with the 
observed $30-50$~ms . 
 
The maximum of the phenomenon can also readily be understood. For short times, 
we have seen that positive velocities along $x$ correspond to one sign for 
the transverse orientation, and conversely. For times greater than 
$2\pi/\sqrt{\omega\delta\Omega}$  ($\delta\Omega$ is the variation of 
$\overline{\Omega}(x)$ between the center and the edge of the cloud), both 
velocity signs become correlated to all spin directions in the transverse 
plane, so that the apparent segregation effect averages out to zero. 
Typical values taken from the experiment of \cite{Cornell} give $\sim 100$~ms for the
maximum of the phenomenon.

When, eventually, two separated species recombine under the effect of the 
restoring force of the trap, the ISRE plays no role anymore. The reason is 
merely that the operator associated with transverse spin is diagonal in the 
position representation (but not in the spin space), so that it can have 
non-zero value only if the wave-packets associated with each internal state 
overlap. In the absence of transverse polarization, the system is equivalent 
to a classical mixture of two gases. 
 
\begin{figure}[ptb] 
\begin{center} 
\includegraphics[ height=2in] {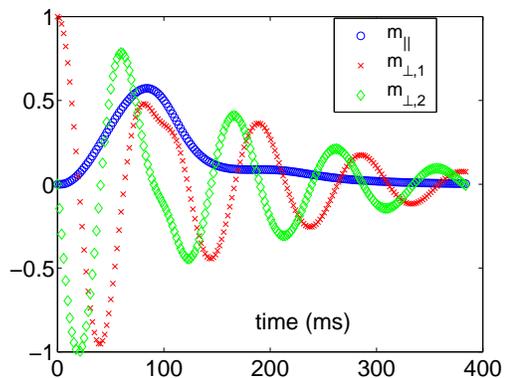} 
\end{center} 
\caption{Time evolution of the spin polarization $\mathbf{m}$ at the center of 
the trap; $m_{\|}$ corresponds to the population difference between the two 
states; $m_{\bot,1}$ and $m_{\bot,2}$ are the two components of the transverse spin polarization. 
The center-to-edge difference in Bohr frequency $\delta\Omega/2\pi$ is 
$12$~Hz.}%
\label{mzt}%
\end{figure} 
 
Numerically, equation (\ref{msimplifie}) can be solved by propagating the 
initial distribution in time with the Lax-Wendroff method (see e.g. 
\cite{numrep}), with parameters taken from ref. \cite{Cornell}. The Bohr 
frequency $\overline{\Omega}(x)$ is taken to be an inverted Gaussian of depth 
$\delta\Omega$ and half-width $x_{T}$. The dimensionless ISRE constant is 
$g_{12}n(0)/\hbar\omega=20$. The relaxation time is $\tau_{M}\sim10$~ms 
\cite{Cornell}, so that $\omega\tau_{M}\simeq0.3$. The time evolution of the 
spin polarization at the center of the trap $\mathbf{m}(0,t)$ is shown in Fig. 
\ref{mzt}, with no adjustable parameter. The longitudinal spin polarization 
 rises as predicted by (\ref{exactresult}), reaches a 
maximum around $90$~ms, and then oscillates and decays to almost zero after 
$300$~ms. Even in the pure Knudsen regime ($\tau_{M}=\infty$), a strong 
maximum of $m_{\|}$ is reached around $100$~ms. Fig. \ref{mzt} also shows how 
the other components of the polarization oscillate and decay. An interesting 
feature  of the experiment of ref.\cite{Cornell} is that 
neither the hydrodynamic nor the collisionless regime are valid along the axis 
as $\omega\tau_{M}\sim1$, so that a study of the full phase-space dynamics is necessary.
 
\begin{figure}[ptb] 
\begin{center} 
\includegraphics[ height=1.5in ] {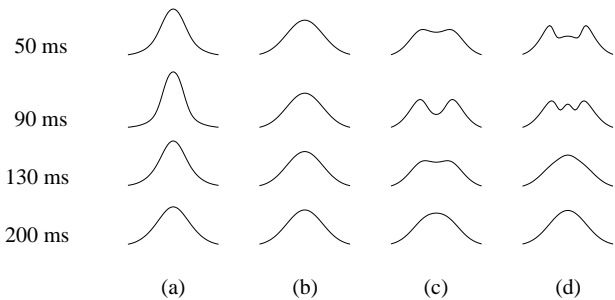} 
\end{center} 
\caption{Evolution of the particle density in state $1$; $\delta\Omega/2\pi$ 
and $g_{12}n(0)/\hbar\omega$ are (a) $-12$~Hz, $20$ (b) $0$~Hz, $20$ (c) 
$12$~Hz, $20$ (d) $30$~Hz, $30$. Note the good agreement with Fig. 3 of 
\cite{Cornell}, including the sign of the effect and ``higher order 
effects''.}%
\label{evol}%
\end{figure}

Fig.~\ref{evol} shows the time evolution of $n_{1}$ as a function of 
$\delta\Omega$, the variation of $\overline{\Omega}(x)$ between the center and 
the edge of the atomic cloud, and of $g_{12}n(0)/\hbar\omega$. When the 
curvature is zero (column (b)), no state separation occurs. Columns (a) 
and (c) show the effect of spin conversion for negative and positive curvature 
$\delta\Omega$ (at the center). For negative curvature, the atoms in state $1$ 
are pulled towards the center of the trap, whereas for positive curvature they 
are expelled from it. Column (d) exhibits what authors of ref. \cite{Cornell} 
call ``higher order effects'', for sufficiently large values of $\delta\Omega$ 
and $g_{12}n(0)/\hbar\omega$. These figures are in good qualitative agreement 
with Fig. 3 in ref. \cite{Cornell}.  

In conclusion, the ISRE plays an important role in the dynamics of cold gases 
with internal states. In the experiment of ref. \cite{Cornell}, this effect 
creates large  longitudinal spin oscillations in a non-hydrodynamic regime. 
Our calculations are also valid for fermions \cite{fermions}, where similar 
effects could be observed, in a case where $g_{11}^{\epsilon}=g_{22}%
^{\epsilon}=0$ and the ISRE changes sign. Another interesting possibility is 
tuning the effect by changing $g_{12}^{\epsilon}$ at a Feshbach resonance 
\cite{Feshbach}. Finally we note that, strictly speaking, our study is limited to 
non-degenerate gases; nevertheless, for non-condensed systems, most of the 
effect of degeneracy can be included by simply replacing the Maxwell-Boltzmann 
distribution by the appropriate quantum distribution \cite{Jeon}, so that no 
dramatic change is expected. 
 
\emph{Note:} while this article was being written, we became aware of the work 
of Oktel and Levitov \cite{Oktel}, who reach conclusions similar to ours. 
Nevertheless, they use an hydrodynamic expression for the evolution of the 
spin current, while here we put more emphasis on the intermediate and Knudsen regimes.\ 
 
\emph{Acknowledgments:} we are grateful to Eric Cornell for explaining the 
results of his group prior to publication, and for several useful discussions. 
We also thank L.S. Levitov for pointing out to us the importance of radial 
averaging \cite{Levitov}. Le LKB est UMR 8552 du CNRS, de l'ENS et de 
l'Universit\'{e} P. et M. Curie.  

\begin{thebibliography}{9}
\bibitem {elastic}All the microscopic processes are purely elastic.
\end{thebibliography}
\end{document}